\documentclass[12pt]{iopart}

\usepackage{iopams} 
\usepackage{graphicx} 
 \renewcommand{\bs}{\boldsymbol}
 \newcommand{\HF}{H_{\rm F}}
 \newcommand{\HA}{H_{\rm A}}

\newcommand{\oL}{\omega}

\newcommand{\ncol}{n^{(\rm col)}}
\newcommand{\ntwoD}{n^{(\rm 2D)}}
\newcommand{\nthreeD}{n^{(\rm 3D)}}
\newcommand{\OD}{{\cal D}}
\newcommand{\ODBL}{{\cal D}_{\rm BL}}

\newcommand{\bss}{\bs s}

\begin{document}

\title[Absorption imaging of a quasi 2D gas]{Absorption imaging of a quasi 2D gas:\\a multiple scattering analysis}

\author{L Chomaz$^1$, L Corman$^{1,2,3}$, T Yefsah$^1$, R Desbuquois$^1$ and J~Dalibard$^1$}
\address{$^1$ Laboratoire Kastler Brossel, CNRS, UPMC, Ecole normale sup\'erieure, \\
24 rue Lhomond, 75005 Paris, France}
\address{$^2$ Ecole polytechnique (member of ParisTech),  91128 Palaiseau cedex, France}
\address{$^3$ Institute for Quantum Electronics, ETH Zurich, 8093 Zurich, Switzerland}
\ead{jean.dalibard@lkb.ens.fr}
\begin{abstract}
Absorption imaging with quasi-resonant laser light is a commonly used technique to probe ultra-cold atomic gases in various geometries. Here we investigate some non-trivial aspects of this method when it is applied to \emph{in situ} diagnosis of a quasi two-dimensional gas. Using Monte Carlo simulations we study the modification of the absorption cross-section of a photon when it undergoes multiple scattering in the gas. We determine the variations of the optical density with various parameters, such as the detuning of the light from the atomic resonance and the thickness of the gas. We compare our results to the known three-dimensional result  (Beer--Lambert law) and outline the specific features of the two-dimensional case.
\end{abstract}
\maketitle

\section{Introduction}

The study of cold atomic gases has recently shed new light on several aspects of quantum many-body physics \cite{Dalfovo:1999,Lewenstein:2007,Bloch:2008,Giorgini:2008}. Most of the measurements in this field of research are based on the determination of the spatial density of the gas \cite{Ketterle:1999b}.  For instance one can use the \emph{in situ} steady-state atomic distribution  in a trapping potential to infer the equation of state of the homogenous gas \cite{Ho:2009}. Another example is the time-of-flight method, in which one measures the spatial density after switching off the trapping potential and allowing for a certain time of ballistic expansion. This gives access to the momentum distribution of the gas, and to the conversion of interaction energy into kinetic energy at the moment of the potential switch-off.

To access the atomic density $n(\bs r)$,  one usually relies on the interaction of the atoms with quasi-resonant laser light. The most common method is absorption imaging, in which the shadow imprinted by the cloud on a low intensity probe beam is imaged on a camera. The simplest modelling of absorption imaging is based on a mean-field approach, in which one assumes that the local value of the electric field driving an atomic dipole at a given location depends only on the average density of scatterers. One can then relate the attenuation of the laser beam to the column atomic density  $\ncol (x,y)=\int n(\bs r)\, dz$ along the line-of-sight $z$. The optical density of the cloud $\OD(x,y)\equiv \ln [I_{\rm in}(x,y)/I_{\rm out}(x,y)]$ is given by the Beer--Lambert law
\begin{equation}
\ODBL(x,y)=\sigma\, \ncol (x,y),
\label{eq:naive_absorption}
\end{equation} 
where $\sigma$ is the photon scattering cross-section, and $I_{\rm in}$ (resp. $I_{\rm out}$) are the incoming (resp. outgoing) intensity of the probe laser in the plane $xy$ perpendicular to the propagation axis. For a closed two-level atomic transition of frequency $\omega_0=ck_0$, $\sigma$ depends on the wavelength $\lambda_0=2\pi /k_0$ associated to this transition and on the detuning $\Delta=\omega-\omega_0$ between the probe light frequency $\omega$ and the atomic frequency: 
\begin{equation}
\sigma=\frac{\sigma_0}{1+\delta^2}, \qquad \sigma_0=\frac{3\lambda_0^2}{2\pi},\qquad 
\delta=\frac{2\Delta}{\Gamma}\ .
\label{eq:sigma0}
\end{equation}
Here $\Gamma$ represents the natural line width of the transition (\emph{i.e.}, $\Gamma^{-1}$ is the natural life time of the excited state of the transition).  Eq. (\ref{eq:sigma0}) assumes that the intensity of the probe beam is much lower than the saturation intensity of the atomic transition. Quasi-resonant absorption imaging is widely used to measure the spatial distribution of atomic gases after a long time-of-flight, when the density has dropped sufficiently so that the  mean-field approximation leading to Eq. (\ref{eq:naive_absorption}) is valid.

One can also use absorption imaging to probe \emph{in situ} samples, at least in the case where $\sigma\, \ncol $ is not very large so that the output intensity is not vanishingly small. This is in particular the case for low dimensional gases. Consider for example a 2D gas, such that the translational degree of freedom along $z$ has been frozen. For a probe beam propagating along this axis, one can transpose the Beer--Lambert law of Eq. (\ref{eq:naive_absorption}) by simply replacing the column density by the surface density $\ntwoD$ of the gas. This 2D Beer--Lambert law can be heuristically justified by treating each atom as a disk of area $\sigma$ that blocks every photon incident on it. In an area $A\gg \sigma$ containing $N=A \ntwoD\gg 1$ randomly placed atoms, the probability that a photon is not blocked by any of the disks is $(1-\sigma/A)^N\approx \exp(-\sigma \ntwoD)$.  

In  a quasi-2D gas there is however an important limitation on the optical densities to which one may apply the Beer-Lambert prediction of Eq.~(\ref{eq:naive_absorption}). Already for $\sigma_0 \ntwoD=1$ the mean interparticle distance is only $0.7\,\lambda_0$ and one may expect that the optical response of an atom strongly depends on the precise location of its neighbours. More precisely the exchange of photons between closely spaced atoms induces a resonant van der Waals interaction that significantly shifts the atomic resonance frequency with respect to its bare value $\omega_0$. The optical density of the gas at resonance may then be reduced with respect to Eq.~(\ref{eq:naive_absorption}), and this was indeed observed in a series of experiments performed with a degenerate $^{87}$Rb gas \cite{Rath:2010,Yefsah:2011}. 

The general subject of the propagation of a light wave in a dense atomic sample, where multiple scattering plays an essential role, has been the subject of numerous experimental and theoretical works (see \emph{e.g.} \cite{Labeyrie:1999,Labeyrie:2003} in the context of cold atoms, and \cite{Akkermans:2007} for a review).
Here we present a quantitative treatment of the collective effects that appear when a weak probe beam interacts with a quasi-2D atomic gas. We consider an ensemble of $N$ atoms at rest with random positions and we investigate the transmission of quasi-resonant light by the atom sheet. We model the resonance transition between the atomic ground ($g$) and excited ($e$) states  by a $J_g=0 \leftrightarrow J_e=1$ transition. We present two equivalent approaches; the first one is based on the calculation of the field radiated by an assembly of $N$ dipoles, where each dipole is driven by an external field plus the field radiated by the $N-1$ other dipoles; the second one uses the standard $T$ matrix formalism of scattering theory. We show that in both cases  the optical density of the medium can be determined by solving the same $3N\times 3N$ linear system. A similar formalism has been previously used for the study of light propagation in small 3D atomic samples, in the presence of multiple scattering (see \emph{e.g.} \cite{Morice:1995,Pinheiro:2004,Gero:2007,Svidzinsky:2008,Akkermans:2008,Sokolov:2009,Scully:2009,Goetschy:2011b}). However its application to quasi-2D samples has (to our knowledge) not yet been investigated, except in the context of Anderson localisation of light  \cite{Pinheiro:2004}.  Our numerical calculations are performed for $N=2048$\,atoms, which is sufficient to reach the `thermodynamic limit' for the range of parameters that is relevant for experiments. We show in particular that even for moderate values of $\sigma_0 \ncol$, the optical density is notably reduced compared to what is expected from the Beer-Lambert law (\emph{e.g.}, more than $20\,\%$ reduction for $\sigma_0 \ncol=1$). We investigate how the absorption line shape is modified by the resonant van der Waals interactions and we also show how the result (\ref{eq:naive_absorption}) is recovered when one increases the thickness of the gas, for a given column density $\ncol$. 

The paper is organised as follows. In section \ref{sec:modelling}, we detail the modelling of the atom-light interaction with the two-level and rotating wave approximations. Then we explain  the principle of the calculation for the absorption of a weak probe beam crossing the atom slab (section \ref{sec:absorption_calculation}). The ensemble of our numerical results are presented in section \ref{sec:results}. Finally  in section \ref{sec:summary} we discuss some limitations to our model and draw some concluding remarks.


\section{Modelling the atom-light interaction}
\label{sec:modelling}

\subsection{The electromagnetic field}

We use the standard description of the quantised electromagnetic field in the Coulomb gauge \cite{cohe92}, and choose  periodic boundary conditions in the cubic-shaped quantisation volume  ${\cal V}=L_x L_yL_z$. We denote $a_{\bs q,\bss}$ the destruction operator of a photon with wave vector $\bs q$ and polarisation $\bss$ ($\bss\bot \bs q$). The Hamiltonian of the quantised field is 
\begin{equation}
\HF=\sum_{\bs q,\bss} \hbar  cq\, a_{\bs q,\bss}^\dagger a_{\bs q,\bss}\ ,
\label{eq:field_H}
\end{equation}
and the transverse electric field operator reads $\bs E(\bs r)=\bs E^{(+)}(\bs r)+\bs E^{(-)}(\bs r)$ with
\begin{equation}
\bs E^{(+)}(\bs r)=i\sum_{\bs q,\bss}\sqrt{\frac{\hbar c q}{2\varepsilon_0 {\cal V}}} \,a_{\bs q,\bss}\,e^{i\bs q\cdot \bs r}\,\bss\ ,
\label{eq:electric_field}
\end{equation}
and $\bs E^{(-)}(\bs r)=\left( \bs E^{(+)}(\bs r) \right)^\dagger$. The wave vectors $\bs q$ are quantised in the volume ${\cal V}$ as $q_i=2\pi n_i /L_i$, $i=x,y,z$, where $n_i$ is a positive or negative integer.

\subsection{The atomic medium}

We consider a collection of $N$ identical atoms at rest in positions $\bs r_j$,  $j=1,\ldots N$. 
We model the atomic resonance transition by a two-level system with a ground state $|g\rangle$ with angular momentum $J_g=0$ and an excited level of angular momentum $J_e=1$. We choose as a basis set for the excited manifold the three Zeeman sublevels $|e_\alpha\rangle$, $\alpha=x,y,z$, where $|e_\alpha\rangle$ is the eigenstate with eigenvalue 0 of the component $J_\alpha$ of the atomic angular momentum operator.
We denote $\hbar \omega_0$ the energy difference between $e$ and $g$. The atomic Hamiltonian is thus (up to a constant)  
\begin{equation}
\HA=\sum_{j=1}^N \sum_{\alpha=x,y,z} \hbar \omega_0 \, |j:e_\alpha\rangle \langle j:e_\alpha| .
\label{eq:atom_H}
\end{equation}

The restriction to a two-level approximation is legitimate if the detuning $\Delta$ between the probe and the atomic frequencies is much smaller than $\omega_0$. The modelling of this transition by a $J_g=0\leftrightarrow J_e=1$ transition leads to a relatively simple algebra. The transitions that are used for absorption imaging in real experiments often involve more Zeeman states  ($J_g=2 \leftrightarrow J_e=3$ for Rb atoms in \cite{Rath:2010,Yefsah:2011}), but are more complex to handle \cite{Jonckheere:2000,Muller:2002} and they are thus out of the scope of this paper. However we believe that the most salient features of multiple scattering and resonant Van der Waals interactions  are captured by our simple level scheme.

\subsection{The atom-light coupling}

We treat the atom-light interaction using the electric dipole approximation (length gauge), which is legitimate since the resonance wavelength of the atoms $\lambda_0$ is much larger than the atomic size. We write the atom-light coupling as:
\begin{equation}
V=-\sum _j \bs D_j \cdot \bs E(\bs r_j) ,
\label{eq:atom_light_coupling0}
\end{equation}
where $\bs D_j$ is the dipole operator for the atom $j$. We will use the rotating wave approximation (RWA), which consists in keeping only the resonant terms in the coupling:
\begin{equation}
V\approx -\sum _j \bs D_j^{(+)} \cdot \bs E^{(+)}(\bs r_j) + \mbox{h.c.}\ ,
\label{eq:atom_light_coupling}
\end{equation}
where h.c. stands for Hermitian conjugate. Here $\bs D_j^{(+)}$ represents the raising part of the dipole operator for atom $j$:
\begin{equation}
\bs D_j^{(+)}= d \sum _{\alpha=x,y,z} |j:e_\alpha\rangle \langle j:g|\, \bs {\hat u_\alpha}\ ,
\label{eq:dipole_operator}
\end{equation}
where $d$ is the electric dipole associated to the $g-e$ transition and $\bs {\hat u_\alpha}$ is a unit vector in the direction $\alpha$.

When a single atom is coupled to the electromagnetic field, this coupling results in the modification of the resonance frequency (Lamb shift) and in the fact that the excited state $e$ acquires a non-zero width $\Gamma$ 
\begin{equation}
\Gamma= \frac{d^2\omega_0^3}{3\pi\varepsilon_0\hbar c^3}\ .
\label{eq:Gamma}
\end{equation}  
For simplicity we will incorporate the Lamb shift in the definition of $\omega_0$. Note that the proper calculation for this shift requires that one goes beyond the two-level and the rotating wave approximations. The linewidth $\Gamma$ on the other hand can be calculated from the above expressions for $V$ using the Fermi golden rule.

The RWA provides a very significant simplification of the treatment of the atom-light coupling, in the sense that the total number of excitations is a conserved quantity. The annihilation (resp. creation) of a photon is always associated with the transition of one of the $N$ atoms from $g$ to $e$ (resp. from $e$ to $g$). This would not be the case if the non-resonant terms of the electric dipole coupling $\bs D_i^{(+)} \cdot \bs E^{(-)}$ and $\bs D_i^{(-)} \cdot \bs E^{(+)}$ were also taken into account. The small parameter associated to the RWA is $\Delta/\omega_0$, which is in practice in the range $10^{-6}-10^{-9}$; the RWA is thus an excellent approximation.

Formally the use of the electric dipole interaction implies to add to the Hamiltonian an additional contact term between the dipoles (see \emph{e.g.} \cite{Cohen:1989,Morice:1995}). This term will play no role in our numerical simulations because we will surround the position of each atom by a small excluded volume, which mimics the short range repulsive interaction between atoms. We checked that the results of our numerical calculations (see Sec. \ref{sec:results}) do not depend on the size of the excluded volume, and we can safely omit the additional contact term in the present work.


\section{Interaction of a  probe laser beam with a dense quasi-2D atomic sample}
\label{sec:absorption_calculation}

We present in this section the general formalism that allows one to calculate the absorption of a quasi-resonant laser beam by a slab of $N$ atoms. We address this question using two different approaches. The first one maps the problem onto the collective behaviour of an assembly of $N$ oscillating dipoles \cite{Morice:1995}. The equation of motion for each dipole is obtained using the Heisenberg picture for the Hamiltonian presented in section \ref{sec:modelling}. It contains two driving  terms, one from the incident probe field and one from the field radiated by all the other dipoles at the location of the dipole under study. The steady-state of this assembly of dipoles is obtained by solving a set of $3N$ linear equations. The second approach uses the standard quantum scattering theory \cite{Messiah:scattering}, which is well suited for perturbative calculations and partial resummations of diagrams.  We suppose that one photon is incident on the atomic medium and we use resummation techniques to take into account the multiple scattering events that can occur before the photon emerges from the medium. The relevant quantity in this approach is the probability amplitude $T_{ ii}$ that the outgoing photon is detected in the same mode as the incident one \cite{Gero:2007,Sokolov:2009}, and we show that $T_{ ii}$ is obtained from the same set of equations as the values of the dipoles in the first approach.
 
\subsection{Wave propagation in an assembly of driven dipoles.}
In this section we assume that the incident field is prepared in a coherent state corresponding to a monochromatic plane wave $E_{\rm L}\,\bs \epsilon\,e^{i(kz-\omega t)}$. We choose the polarization $\bs \epsilon$ to be linear and parallel to the $x$ axis  ($\bs \epsilon=\bs{\hat u_x}$). Since we consider a $J_g=0 \leftrightarrow J_e=1$ transition, this choice does not play a significant role and we checked that we recover essentially the same results with a circular polarisation. Note that the situation would be different for an atomic transition with larger $J_g$ and $J_e$ since optical pumping processes would then depend crucially on the polarisation of the probe laser.

The amplitude $E_{\rm L}$  is  supposed to be small enough that the steady-state populations of the excited states $e_{j,\alpha}$ are small compared to unity. This ensures that the response of each atomic dipole is linear in $E_{\rm L}$;  this approximation is valid when the Rabi frequency $dE_{\rm L}/\hbar$ is  small compared to the natural width $\Gamma$ or the detuning $\Delta$. 

Using the atom-light coupling (\ref{eq:atom_light_coupling0}), the equations of motion for the annihilation operators $a_{\bs q,\bss}$ in the Heisenberg picture read:
\begin{equation}
\dot a_{\bs q,\bss}(t)=-i \,cq\, a_{\bs q,\bss}(t)+\sqrt{\frac{cq}{2\hbar \varepsilon_0 {\cal V}}} \sum_{j'} \bss^*\cdot \bs D_{j'}(t)\,e^{-i\bs q\cdot \bs r_{j'}}\, .
\label{eq:motion_a}
\end{equation}
This equation can be integrated between the initial time $t_0$ and the time $t$, and the result can be injected in the expression for the transverse  field to provide its value at any point $\bs r$:
\begin{eqnarray}
E_{\alpha}(\bs r,t)&=&E_{{\rm free},\alpha}(\bs r,t)+\sum_{j',\alpha'}\sum_{\bs q,\bss} \int_0^{t-t_0}d\tau\; \frac{cq}{2\varepsilon_0 {\cal V}}
\nonumber\\
&&\,\left[ i D_{j',\alpha'}(t-\tau)\,e^{i\bs q\cdot (\bs r-\bs r_{j'})-icq \tau}\,s_\alpha s^*_{\alpha'} + \mbox{h.c.}\right]\ ,
\label{eq:field_Heisenberg}
\end{eqnarray}
where $\bs E_{{\rm free}}$ stands for the value obtained in the absence of atoms. We now take the quantum average of this set of equations. In  the steady-state regime the expectation value of the dipole operator $\bs D_j (t)$ can be written  $\bs d_j e^{-i\omega t} + \mbox{c.c.}$, and the average of $\bs E_{{\rm free}}(\bs r,t)$ is the incident field $E_{\rm L}\,\bs \epsilon\,e^{i(kz-\omega t)}+\mbox{c.c.}$\ . We denote the average value of the transverse field operator in $\bs r$ as $\langle \bs E(\bs r,t)\rangle =\bar {\bs E} (\bs r)\,e^{-i\omega t}+\mbox{c.c.}$, and we obtain 
after some algebra (see \emph{e.g.} \cite{Morice:1995,Morice:1995b})
\begin{equation}
\bar{E}_{\alpha}(\bs r)=E_{{\rm L}}\,\epsilon_{\alpha}\,e^{i kz}+\frac{k^3}{6\pi\varepsilon_0}\sum_{j',\alpha'}g_{\alpha,\alpha'}(\bs u_{j'})\,d_{j',\alpha'}\ ,
\label{eq:field_average}
\end{equation}
where we set $\bs u_j=k (\bs r-\bs r_j)$ (with $k\approx k_0$), 
\begin{equation}
g_{\alpha,\alpha'}(\bs u)=\delta_{\alpha,\alpha'}h_1(u)+\frac{u_\alpha u_{\alpha'}}{u^2}h_2(u),
\label{eq:g}
\end{equation}
and
\begin{equation}
h_1(u)=\frac{3}{2}\,\frac{e^{iu}}{u^3}(u^2+iu-1),\qquad
h_2(u)=\frac{3}{2}\,\frac{e^{iu}}{u^3}(-u^2-3iu+3) .
\label{eq:h1h2}
\end{equation}
The function $g_{\alpha,\alpha'}(k\bs r)$ is  identical to the one appearing in classical electrodynamics \cite{Jackson:book}, when calculating the field radiated in $\bs r$ by a dipole located at the origin. 

We proceed similarly for the equations of motion for the dipole operators $\bs D_j^{(-)}$ and take their average value in steady-state. The result can be put in the form \cite{Morice:1995}
\begin{equation}
 (\delta+i) d_{j,\alpha} + \sum_{j'\neq j,\;\alpha'} g_{\alpha,\alpha'}(\bs u_{jj'}) d_{j',\alpha'}=  - \frac{6\pi\varepsilon_0}{k^3}\,  E_{{\rm L}}\,\epsilon_{\alpha}\,e^{i kz_j}\ ,
\label{eq:sys_dipoles}
\end{equation}
where the reduced detunig $\delta=2\Delta/\Gamma$ has been defined in Eq. (\ref{eq:sigma0}) and $\bs u_{j,j'}=k(\bs r_{j'}-\bs r_j)$. This can be written with matrix notation
\begin{equation}
[M] |X\rangle=|Y\rangle
\label{eq:sys_matrix}
\end{equation}
where the $3N$ vectors $|X\rangle$ and $|Y\rangle$ are defined by
\begin{equation}
X_{j,\alpha}=-\frac{k^3}{6\pi \epsilon_0 E_L}d_{j,\alpha},
\qquad
Y_{j,\alpha}=\epsilon_{\alpha}e^{ik z_{j}}\ ,
\label{eq:def_X_Y}
\end{equation} 
and where the complex symmetric matrix $[M]$ has its diagonal coefficients equal to $\delta +i$ and its off-diagonal coefficients (for $j\neq j'$) given by $g_{\alpha,\alpha'}(\bs u_{jj'})$. This matrix belongs to the general class of Euclidean matrices \cite{Mezard:1999}, for which the $(i,j)$ element can be written as a function $F(\bs r_i,\bs r_j)$ of points $\bs r_i$ in the Euclidean space. The spectral properties of these matrices for a random distribution of the $\bs r_i$'s (as it will be the case in this work, see Sec. \ref{sec:results}) have been studied in \cite{Mezard:1999,Rusek:2000,Skipetrov:2011,Goetschy:2011}.

Eq. (\ref{eq:sys_dipoles}) has a simple physical interpretation: in steady-state each dipole $\bs d_j$ is driven by the sum of the incident field $ E_{\rm L}$ and the field radiated by all the other dipoles. This set of $3N$ equations was first introduced by L. L. Foldy in \cite{Foldy:1945} who named it, together with Eq. (\ref{eq:field_average}), ``the fundamental equations of multiple scattering". Indeed for a given incident field, the solution of  (\ref{eq:sys_matrix}) provides the value of each dipole $\bs d_j$, which can then be injected in (\ref{eq:field_average}) to obtain the value of the total field at any point in space. 

\subsection{Absorption signal}
\label{subsec:absorption_signal}

\begin{figure}[t]
\begin{center}
\includegraphics{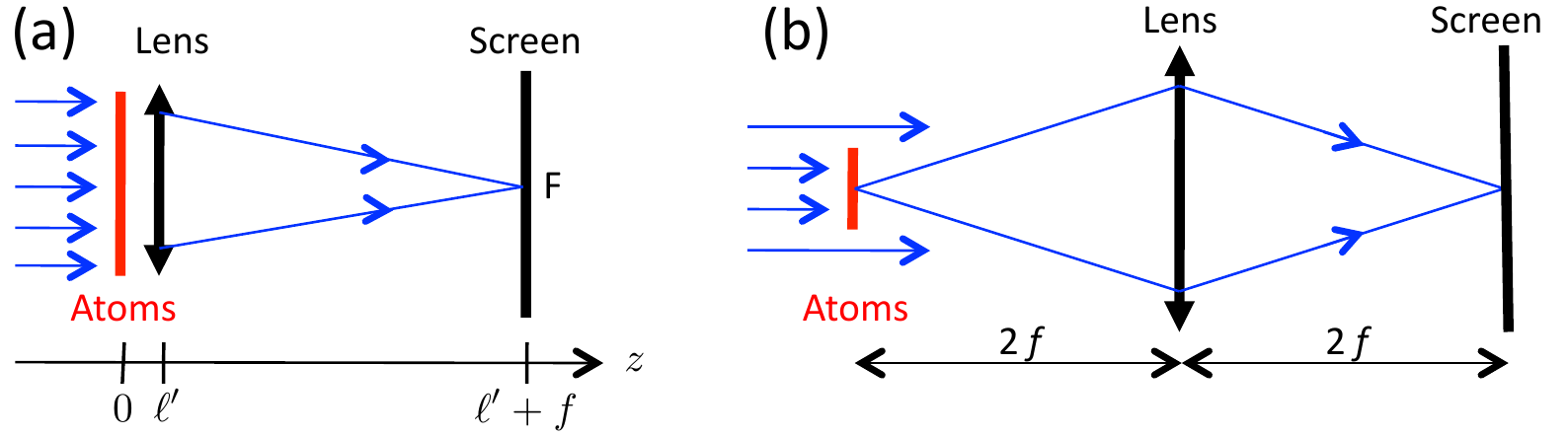}
\caption{Two possible setups for measuring the absorption of an incident probe beam by a slab of atoms using a lens of focal $f$. a) Global probe. b) Local probe.}
\label{fig:experiment}
\end{center}
\end{figure}

From the expression of the average value of the dipoles  we now extract the absorption coefficient of the probe beam and the optical density of the gas. We suppose that the $N$ atoms are uniformly spread in a cylinder of radius $R$ along the $z$ axis and  located between $z=-\ell/2$ and $z=\ell/2$.  We can consider two experimental setups  to address this problem. The first one, represented in Fig. \ref{fig:experiment}a, consists in measuring after the atomic sample the total light intensity with the same momentum $\bs k=k \bs{\hat u_z}$  as the incident probe beam. This can be achieved by placing a lens with the same size as the atomic sample, in the plane $z=\ell'>\ell/2$ just after the sample. The light field at the focal point of the lens $F$ gives the desired attenuation coefficient. We refer to this method as `global', since the field $\bs E(F)$  provides information over the whole atomic cloud. One can also use the setup sketched in Fig. \ref{fig:experiment}b, which forms an image of the atom slab on a camera and provides a `local' measurement of the absorption coefficient.  In real experiments  local measurements are often favored because trapped atomic sample are non homogeneous and it is desirable to access the spatial distribution of the particles. However for our geometry with a uniform density of scatterers, spatial information on the absorption of the probe beam is not  relevant. Therefore we only present the formalism for global measurements, which is simpler to derive and leads to slightly more general expressions.   We checked numerically that we obtained very similar results when we modelled  the local procedure.

We assume that the lens in Fig. \ref{fig:experiment}a operates in the paraxial regime, \emph{i.e.}, its focal length $f$ is much larger than its radius $R$. We relate the field at the image focal point of the lens to the field in the plane $z=\ell'$ just before the lens:
\begin{equation}
\bs E(F) =-\frac{i e^{ikf}}{\lambda_0 f}  \int_{\cal L} \bs E (x,y,\ell')\;dx\,dy ,
\label{eq:lens_global}
\end{equation}
where the integral runs over the lens area. Since the incident probe beam is supposed to be linearly polarised along $x$, we calculate the $x$ component of the field in $F$. Plugging the value of the field given in Eqs. (\ref{eq:field_average},\ref{eq:def_X_Y})  we obtain the transmission coefficient
\begin{equation}
{\cal T}\equiv \frac{\left. E_x(F)\right|_{\rm with\ atoms}}{\left.E_x(F)\right|_{\rm no\ atom}} =  1-\frac{e^{-ik\ell'}}{\pi R^2}\sum_{j,\alpha} X_{j,\alpha}\int_{\cal L}g_{x,\alpha}[k(\bs r-\bs r_j)]\;dx\,dy\ . 
\label{eq:Ex_sur_EL}
\end{equation}
This result can be simplified in the limit of a large lens by using an approximated value for the integral appearing in (\ref{eq:Ex_sur_EL}). We suppose that $k\ell' \gg1$  so that the dominant part in $g_{x,\alpha}$ is the $e^{iu}/u$ contribution to $h_1$. More precisely the domain in the lens plane contributing to the integral for the dipole $j$ is essentially a disk of radius $\sqrt {\lambda (\ell'-z_j)}\sim \sqrt{\lambda \ell'}$ centered on $(x_j,y_j)$. When this small disk is entirely included in the lens aperture, \emph{i.e.}, the larger disk of radius $R$ centered on $x=y=0$, we  obtain
\begin{equation}
 \int_{\cal L}g_{x,\alpha}[k(\bs r-\bs r_j)]\;dx\,dy \approx \frac{3i\pi}{k^2}\delta_{x,\alpha}e^{ik(\ell'-z_j)}\ .
\label{eq:approx_integrale}
\end{equation}
We use the result (\ref{eq:approx_integrale}) for all atoms, which amounts to neglect edge effects for the dipoles located at the border of the lens, and we obtain:
\begin{equation}
{\cal T}= 1-\frac{i}{2}\sigma_0 \ncol \Pi ,
\label{eq:relation_T_Pi}
\end{equation}
with $\ncol=N/\pi R^2$ and where the coefficient $\Pi$ is defined by
\begin{equation}
 \Pi=\frac{1}{N}\sum_j X_{j,x}e^{-ikz_j} .
\label{eq:def_Pi}
\end{equation}
This coefficient captures the whole physics of multiple scattering and resonant van der Waals interactions among the $N$ atoms. Indeed one takes into account all possible couplings between the dipoles when solving the $3N\times 3N$ system $[M] |X\rangle=|Y\rangle$. Once ${\cal T}$ is known the optical density is obtained from
\begin{equation}
\OD\equiv \ln \left|{\cal T}\right|^{-2} .
\label{eq:def_OD_vs T}
\end{equation}

As an example, consider the limit of a very sparse sample where multiple scattering does not play a significant role ($\sigma_0 \ncol\ll 1$). All non-diagonal matrix elements in $[M]$ are then negligible and  $[M]$ is simply the identity matrix, times $ i+\delta$. Each $X_{j,x}$ solution of the system (\ref{eq:sys_matrix}) is equal to $e^{ikz_j}/(i+\delta)$, and we obtain as expected:
\begin{equation}
\sigma_0 \ncol\ll 1:\quad {\cal T}\approx 1-\frac{1}{2(1-i\delta)}\sigma_0 \ncol\ ,\quad {\cal D}\approx \frac{\sigma_0 \ncol}{1+\delta^2} .
\label{eq:sparse}
\end{equation}

\subsection{Light absorption as a quantum scattering process}
\label{subsec:scatt_approach}

In order to study the attenuation of a weak probe beam propagating along the $z$ axis when it crosses the atomic medium, we can also use quantum scattering theory. The  Hamiltonian of the problem is
\begin{equation}
H=H_0+V\ ,\qquad H_0=\HA+\HF,
\label{eq:H_H0}
\end{equation}
and we consider the initial state where all atoms are in their ground state and where a single photon of wave vector $\bs k=k\bs{\hat u_z}$ and polarisation $\bs \epsilon=\bs {\hat u_x}$ is incident on the atomic medium
\begin{equation}
|\Psi_{\rm i}\rangle =|{\cal G}\rangle \otimes |\bs k, \bs \epsilon\rangle ,
\label{eq:Psi_i}
\end{equation}
with $|{\cal G}\rangle\equiv |1:g,\ 2:g,\ \ldots, N:g\rangle$.  The state  $|\Psi_{\rm i}\rangle$ is an eigenstate of $H_0$ with energy $\hbar \omega$. The interaction of the photon with the atomic medium, described by the coupling $V$, can be viewed as a collision process during which an arbitrary number of elementary scattering events can take place. Each event starts from a state $|{\cal G}\rangle \otimes |\bs q, \bss\rangle$ and
corresponds to: 
\begin{itemize}
 \item[(i)]
The absorption of the photon in mode $\bs q, \bss$ by atom $j$, which jumps from its ground state $|j:g\rangle$ to one of its excited states $|j:e_\alpha\rangle$. The state of the system is then
\begin{equation}
 |{\cal E}_{j,\alpha}\rangle=|1:g,\ \ldots ,j:e_\alpha,\ldots,N:g\rangle \otimes |\mbox{vac}\rangle,
\label{eq:def_E_j}
\end{equation}
where $|\mbox{vac}\rangle$ stands for the vacuum state of the electromagnetic field.
The subspace spanned by the states $|{\cal E}_{j,\alpha}\rangle$ has dimension $3N$.
 \item[(ii)]
The emission of a photon in the mode $(\bs q',\bss')$ by  atom $j$, which falls back into its ground state.
\end{itemize}
Finally a photon emerges from the atomic sample, and  we want to determine  the probability amplitude to find this photon in the same mode $|\bs k, \bs \epsilon\rangle$ as the initial one. 

The $T$ matrix defined as
\begin{equation}
T(E)=V+V\frac{1}{E-H+i0_+}V\ ,
\label{eq:def_T}
\end{equation}
where $0_+$ is a small positive number that tends to zero at the end of the calculation, 
provides a convenient tool to calculate this probability amplitude. Generally 
\begin{equation}
T_{ if}=\langle \Psi_{ f}|T(E_i)|\Psi_{ i}\rangle
\label{eq:T_if}
\end{equation}  
gives the probability amplitude to find the system in the final state $|\Psi_{ f}\rangle$ after the scattering process. The states $|\Psi_{ i}\rangle$ and $|\Psi_{ f}\rangle$ are eigenstates of the unperturbed Hamiltonian $H_0$, with energy $E_i$. Here we are interested in the element $T_{ ii}$ of the $T$ matrix, corresponding to the choice  $|\Psi_{ f}\rangle=|\Psi_{ i}\rangle$. Using the definition (\ref{eq:def_T}) we find
\begin{equation}
T_{ ii}=\frac{\hbar \oL d^2}{2\varepsilon_0 {\cal V}}\sum_{j,j'} e^{ik (z_{j}-z_j')} \;\langle {\cal E}_{j',x}|
\frac{1}{\hbar \oL-H+i0_+}
 |{\cal E}_{j,x}\rangle .
\label{eq:Tii}
\end{equation} 

We now have to calculate the $(3N)\times(3N)$ matrix elements of the operator $1/(z-H)$, with $z=\hbar \oL+i0_+$, entering into (\ref{eq:Tii}). We introduce the two orthogonal projectors $P$ and $Q$, where $P$ projects on the subspace with zero photon, and $Q$ projects  on the orthogonal subspace. We thus have
\begin{eqnarray}
P |{\cal E}_{j,\alpha}\rangle= |{\cal E}_{j,\alpha}\rangle
\qquad \qquad
P|{\cal G}\rangle \otimes |\bs k, \bs \epsilon\rangle=0, 
\label{eq:actionP}
\\
Q |{\cal E}_{j,\alpha}\rangle= 0 \qquad \qquad \quad \ \ \,
Q|{\cal G}\rangle \otimes |\bs k, \bs \epsilon\rangle=|{\cal G}\rangle \otimes |\bs k, \bs \epsilon\rangle.\qquad
\label{eq:actionQ}
\end{eqnarray}
We define the displacement operator  
\begin{equation}
R(z)=V+V\frac{Q}{z-QH_0Q-QVQ}V 
\label{eq:def_Rz}
\end{equation}
and use the general result \cite{cohe92}
\begin{equation}
P\frac{1}{z-H}P=\frac{P}{z-H_{\rm eff}}\ ,
\label{eq:resolvant}
\end{equation}
where the effective Hamiltonian $H_{\rm eff}$ is
\begin{equation}
H_{\rm eff}=P\left(H_0+R(z) \right)P .
\label{eq:effectiveH}
\end{equation}

For the following calculations, it is convenient to introduce the dimensionless matrix $[M]$ proportional to the denominator of the right hand side of (\ref{eq:resolvant}):
\begin{equation}
[M]_{(j',\alpha'),(j,\alpha)}= \frac{2}{\hbar \Gamma}\langle {\cal E}_{j',\alpha'}|
 z-H_{\rm eff}|{\cal E}_{j,\alpha}\rangle .
\label{eq:def_M_resolvant}
\end{equation}
It is straightforward to check\footnote{As for the derivation leading from Eq. (\ref{eq:motion_a})
to Eq. (\ref{eq:field_average}), one must take into account the non-resonant terms that are usually dropped in the RWA, in order to ensure the proper convergence of the sum (\ref{eq:matrixlementR}) and obtain the tensor $g_{\alpha\alpha'}$.} that for $z\to \hbar \omega$ this matrix coincides with the symmetric matrix appearing in (\ref{eq:sys_matrix}). Indeed the matrix elements of $R(z) $  are 
\begin{equation}
\langle {\cal E}_{j',\alpha'}|
R(z)  |{\cal E}_{j,\alpha}\rangle=\frac{\hbar  d^2}{2\varepsilon_0 {\cal V}}\sum_{\bs q,\bss}
cq\,
 s_{\alpha}^* s_{\alpha'}\;
 \frac{e^{i\bs q\cdot (\bs r_{j'}-\bs r_j)}}{z-\hbar \omega},
\label{eq:matrixlementR}
\end{equation}
which can be calculated explicitly. For $j=j'$, the real part of this expression is the Lamb shift that we reincorporate in the definition of $\omega_0$,  and its imaginary part reads:
\begin{equation}
\langle {\cal E}_{j,\alpha'}|
R(z)  |{\cal E}_{j,\alpha}\rangle=-i\frac{\hbar \Gamma}{2}\,\delta_{\alpha,\alpha'}\ .
\label{eq:diagRz}
\end{equation}
For $j\neq j'$, the sum over $(\bs q,\bss)$ appearing in (\ref{eq:matrixlementR}) is the propagator of a photon from an atom in $\bs r_j$ in internal state $|e_\alpha\rangle$, to another atom in $\bs r_{j'}$  in internal state $|e_{\alpha'}\rangle$. This is nothing but (up to a multiplicative coefficient) the expression that we already introduced for the field radiated in  $\bs r_{j'}$ by a dipole located in $\bs r_{j}$: 
\begin{equation}
\langle {\cal E}_{j',\alpha'}|
R(z)  |{\cal E}_{j,\alpha}\rangle=-\frac{\hbar \Gamma}{2} g_{\alpha,\alpha'}(\bs u_{j,j'}) , 
\label{eq:propagator}
\end{equation}
where the tensor $g_{\alpha,\alpha'}$ is defined in Eqs. (\ref{eq:g}-\ref{eq:h1h2}).

Suppose now that the atoms are uniformly distributed over the transverse area $L_xL_y$ of the quantisation volume. We set $\ncol=N/(L_xL_y)$ and we rewrite the expression (\ref{eq:Tii}) of the desired matrix element $T_{ ii}$ as
\begin{equation}
\frac{ T_{ ii}L_z}{\hbar c}=\frac{1}{2N}\sigma_0 \ncol \;\sum_{j,j'}e^{i k (z_{j}-z_{j'})} \;[M^{-1}]_{(j,x),(j',x)}=\frac{1}{2}\sigma_0 \ncol \Pi\ ,
\label{eq:Tii_and_M}
\end{equation} 
where the coefficient $\Pi$ has been defined in (\ref{eq:def_Pi}). The result (\ref{eq:Tii_and_M}) combined with (\ref{eq:relation_T_Pi}) leads to
\begin{equation}
{\cal T}=1-i\frac{ T_{ ii}L_z}{\hbar c}\ ,
\label{eq:optical_theorem}
\end{equation}
which constitutes the `optical theorem' for our slab geometry, since it relates the attenuation of the probe beam ${\cal T}$ to the forward scattering amplitude $T_{ii}$.

The emergence of resonant van der Waals interactions is straightforward in this approach. Let us consider for simplicity the case where only $N=2$ atoms are present. The effective Hamiltonian $H_{\rm eff}$ is a $6\times 6$ matrix that can be easily diagonalized  and its eigenvectors, with one atom in $|e\rangle$ and one  in $|g\rangle$,  form in this particular case  an orthogonal basis, although $H_{\rm eff}$ is non-Hermitian \cite{Stephen:1964,Hutchinson:1964}. For a short distance $r$ between the atoms ($kr\ll 1$), the leading term in $h_1(u)$ and $h_2(u)$ is $u^{-3}$ and the energies (real parts of the eigenvalues) of the six eigenstates vary as $\sim \pm \hbar \Gamma/(kr)^3$ (resonant dipole-dipole interaction). The imaginary parts of the eigenvalues, which give the inverse of the radiative lifetime of the states, tend either to $\Gamma$ or $0$ when $r\to 0$, which correspond to the superradiant and subradiant states for a pair of atoms, respectively \cite{Dicke:1954}. 

For $N>2$ the eigenvectors of the non-Hermitian Euclidean matrix $H_{\rm eff}$ are in general non orthogonal, which complicates the use of standard techniques of spectral theory in this context  \cite{Skipetrov:2011,Goetschy:2011}.  More precisely, one could think of solving the linear system (\ref{eq:sys_matrix}), or equivalently calculating $T_{ii}$ in Eq. (\ref{eq:Tii}), by using the expansion of the column vector $|Y\rangle$ defined in Eq. (\ref{eq:def_X_Y}) on the left ($|\alpha_j\rangle$) and right ($\langle \beta_j|$) eigenvectors of $H_{\rm eff}$. Then one could inject this expansion in the general expression of the matrix element $T_{ii}$, to express it as a sum of the contributions of the various eigenvalues of $H_{\rm eff}$. However the physical discussion based on this approach is made difficult by the fact that since $H_{\rm eff}$ is non-Hermitian,  the $\{|\alpha_j\rangle\} $ and the $\{|\beta_j\rangle\}$ bases do not coincide. Hence the weight $\langle \beta_j|Y\rangle \langle Y|\alpha_j\rangle$ of a given eigenvalue in the sum providing the value of $T_{ii}$ is not a positive number, and this complicates the interpretation of the result.

\subsection{Beyond the sparse sample case: 3D vs. 2D}

For a sparse sample, we already calculated the optical density at  first order in density (Eq. (\ref{eq:sparse})) and the result is identical for a strictly 2D gas and a thick one. The approach based on quantum scattering theory is well suited to go beyond this first order approximation and look for differences between the 2D and 3D cases. The basis of the calculation is the series expansion of Eq. (\ref{eq:resolvant}), which gives
\begin{equation}
P\frac{1}{z-H}P = \frac{P}{z-H_0} +\sum_{n=1}^{\infty} \frac{P}{z-H_0}
\left( PR(z)P \frac{1}{z-H_0}\right)^n .
\label{eq:expand_resolvant}
\end{equation}
Consider the case of a resonant probe $\delta=0$ for simplicity. The result ${\cal T}\approx 1-\sigma_0 \ncol/2$ obtained for a sparse sample in Eq. (\ref{eq:sparse}) corresponds to the first term [$P/(z-H_0)$] of this expansion. Here we investigate the next order term and explain why one can still recover the Beer-Lambert law for a thick (3D) gas, but not for a 2D sample.

\paragraph{Double scattering diagrams for a thick sample ($k\ell\gg 1$).}
\label{subsec:double_scattering}

We start our study by adding the first term ($n=1$) in the expansion (\ref{eq:expand_resolvant}) to the zero-th order term already taken into account in Eq. (\ref{eq:sparse}). This amounts to take into account the diagrams where the incident photon is scattered on a single atom, and those where the photon `bounces' on two atoms before leaving the atomic sample. Injecting the first two terms of the expansion (\ref{eq:expand_resolvant}) into (\ref{eq:Tii_and_M}), we obtain
\begin{equation}
\frac{T_{ ii}L_z}{\hbar c}=\frac{1}{2}\sigma_0 \ncol \left[
-i
+ \frac{1}{N}
\sum_j \sum_{j'\neq j} e^{ik(z_j-z_{j'})}g_{xx}(\bs u_{jj'})\right] .
\label{eq:Tii_ordre_deux}
\end{equation}
We now have to average this result on the positions of the atoms $j$ and $j'$. There are $N(N-1)\approx N^2$ couples $(j,j')$. Assuming that the gas is dilute so that the average distance between two atoms (in particular $|z_j-z_{j'}|$) is much larger than $k^{-1}$, the leading term in $g_{xx}$ is the $e^{iu}/u$ contribution of $h_1 (u)$ in Eqs. (\ref{eq:g})-(\ref{eq:h1h2}). We thus arrive at
\begin{equation}
\frac{T_{ ii}L_z}{\hbar c}=\frac{1}{2}\sigma_0 \ncol \left[-i+ \frac{3N}{2k}
\langle e^{ik(z-z')} \frac{e^{ik |\bs r-\bs r'|}}{|\bs r-\bs r'|} \rangle \right],
\label{eq:Tii_ordre_deux_bis}
\end{equation}
where the average is taken over the positions $\bs r$ and $\bs r'$ of two atoms. We first calculate the average over the $xy$ coordinates and we get (cf. Eq. (\ref{eq:approx_integrale}))
\begin{equation}
\frac{T_{ ii}L_z}{\hbar c}=\frac{1}{2}\sigma_0 \ncol \left[-i + \frac{i}{2}\sigma_0 \ncol
\langle e^{ik (z-z')} e^{ik |z-z'|}\rangle \right] .
\label{eq:Tii_average_ordre_deux}
\end{equation}
For a thick gas ($k\ell\gg 1$) the bracket in this expression has an average value of $\approx 1/2$. Indeed the function to be averaged is equal to 1 if $z<z'$, which occurs in half of the cases, and it oscillates and averages to zero in the other half of the cases, where $z>z'$.
We thus obtain the approximate value of the transmission coefficient: 
\begin{equation}
k\ell \gg 1: \quad {\cal T}=1-i\frac{T_{ ii}L_z}{\hbar c}\approx 1-\frac{1}{2}\sigma_0 \ncol + \frac{1}{8} \left( \sigma_0 \ncol \right)^2 ,
\label{eq:transmission_ordre_deux}
\end{equation}
 where we recognize the first three terms of the power series expansion of ${\cal T}=\exp(-\sigma_0 \ncol/2)$, corresponding to the optical density 
 $\OD=\sigma_0 \ncol$.
  
\paragraph{Double scattering diagrams for a 2D gas ($\ell=0$).}
When all atoms are sitting in the same plane, the evaluation of the second order term (and the subsequent ones) in the expansion of $T_{ ii}$ in powers of the density is modified with respect to the 3D case. The calculation starts as above and the second term in the bracket of Eq. (\ref{eq:Tii_ordre_deux}) can now be written 
\begin{equation}
\frac{1}{N}
\sum_j \sum_{j'\neq j} g_{xx}(u_{jj'})=\ntwoD \int g_{xx}(\bs u)\;d^2u\ .
\label{eq:Tii_ordre_deux_twoD_case}
\end{equation}
If we keep only the terms varying as $e^{iu}/u$ in $h_1$ and $h_2$ (Eq. (\ref{eq:h1h2})), we can calculate analytically the integral in (\ref{eq:Tii_ordre_deux_twoD_case}) and find the same result  
as in 3D, \emph{i.e.}, $i\sigma_0\ntwoD/4$. If this was the only contribution to (\ref{eq:Tii_ordre_deux_twoD_case}), it would lead to the Beer--Lambert law also in 2D, at least at second order in density. However one can check that a significant contribution to the integral in (\ref{eq:Tii_ordre_deux_twoD_case}) comes from the region $u=k r < 1$. In this region, it is not legitimate to keep only the term in $e^{ikr}/kr$ in $h_1$, $h_2$, since the terms in $e^{ikr}/(kr)^3$, corresponding to the short range resonant van der Waals interaction, are actually dominant. Therefore the expansion of the transmission coefficient ${\cal T}$ in powers of the density differs from (\ref{eq:transmission_ordre_deux}), and one cannot recover the Beer--Lambert law at second order in density. Calculating analytically corrections to this law could be done following the procedure of \cite{Morice:1995}. Here we will use a numerical method to determine the deviation with respect to the Beer--Lambert law (see
section \ref{subsec:OD_vs_ODBL}).

\paragraph{Remark.} For a 3D gas there are also corrections to the second term in Eq. (\ref{eq:Tii_average_ordre_deux}) due the $1/r^3$ contributions to $h_1$ and $h_2$. However these corrections have a different scaling with the density and can be made negligible. More precisely their order of magnitude is $\sim \nthreeD k^{-3}$, to be compared with the value $\sim \ncol k^{-2}$ of the second term in Eq.\,(\ref{eq:Tii_average_ordre_deux}). Therefore one can have simultaneously $\nthreeD k^{-3} \ll 1$ and $\ncol k^{-2}\gtrsim 1$, if the thickness $\ell$ of the gas along $z$ is $\gg 1/k$.


\section{Absorption of light by a slab of atoms}
\label{sec:results}

In order to study quantitatively the optical response of a quasi-2D gas, we have performed a Monte Carlo calculation of the transmission factor ${\cal T}$ given  in Eq. (\ref{eq:relation_T_Pi}), and of the related optical density $\OD=\ln|{\cal T}|^{-2}$. We start our calculation by randomly drawing the positions of the $N$ atoms, we then solve numerically the $3N\times 3N$ linear system (\ref{eq:sys_matrix}), and finally inject the result for the $N$ dipoles in the expression of ${\cal T}$. 

The atoms are uniformly distributed in a cylinder of axis $z$, with a radius $R$ and a thickness $\ell$. The largest spatial densities considered in this work correspond to a mean inter-particle distance $\approx k^{-1}$. Around each atom we choose a small excluded volume with a linear size $a=0.01\,k^{-1}$. We varied $a$ by a factor 10 around this value and checked that our results were essentially unchanged. Apart from this excluded volume we do not include any correlation between the positions of the atoms. This choice is justified physically by the fact that,  in the case of large phase space densities which motivates our study, the density fluctuations in a 2D Bose gas are strongly reduced and the two-body correlation function $g_2(\bs r,\bs r')$ is such that $g_2(\bs r,\bs r)\approx 1$ \cite{Prokofev:2001}.

In this section we  first determine the value of $N$ that is needed to reach the `thermodynamic limit' for our problem: for a given thickness $\ell$, $\OD$ should not be an independent function of the number of atoms $N$ and the disk radius $R$, but should depend only of the ratio $N/\pi R^2=\ncol$. We will see that this imposes to use relatively large number of atoms, typically $N > 1000$, for the largest spatial densities considered here. All subsequent calculations are performed with $N=2048$. We then study the dependence of $\OD$ with the various parameters of the problem: the column density $\ncol$, the thickness of the gas $\ell$, and the detuning $\Delta$. In particular we show  that for a given $\ncol$ we recover the 3D result (\ref{eq:naive_absorption}) when the thickness $\ell$ is chosen sufficiently large.

\subsection{Reaching the `thermodynamic limit'}

\begin{figure}[t]
\begin{center}
\includegraphics{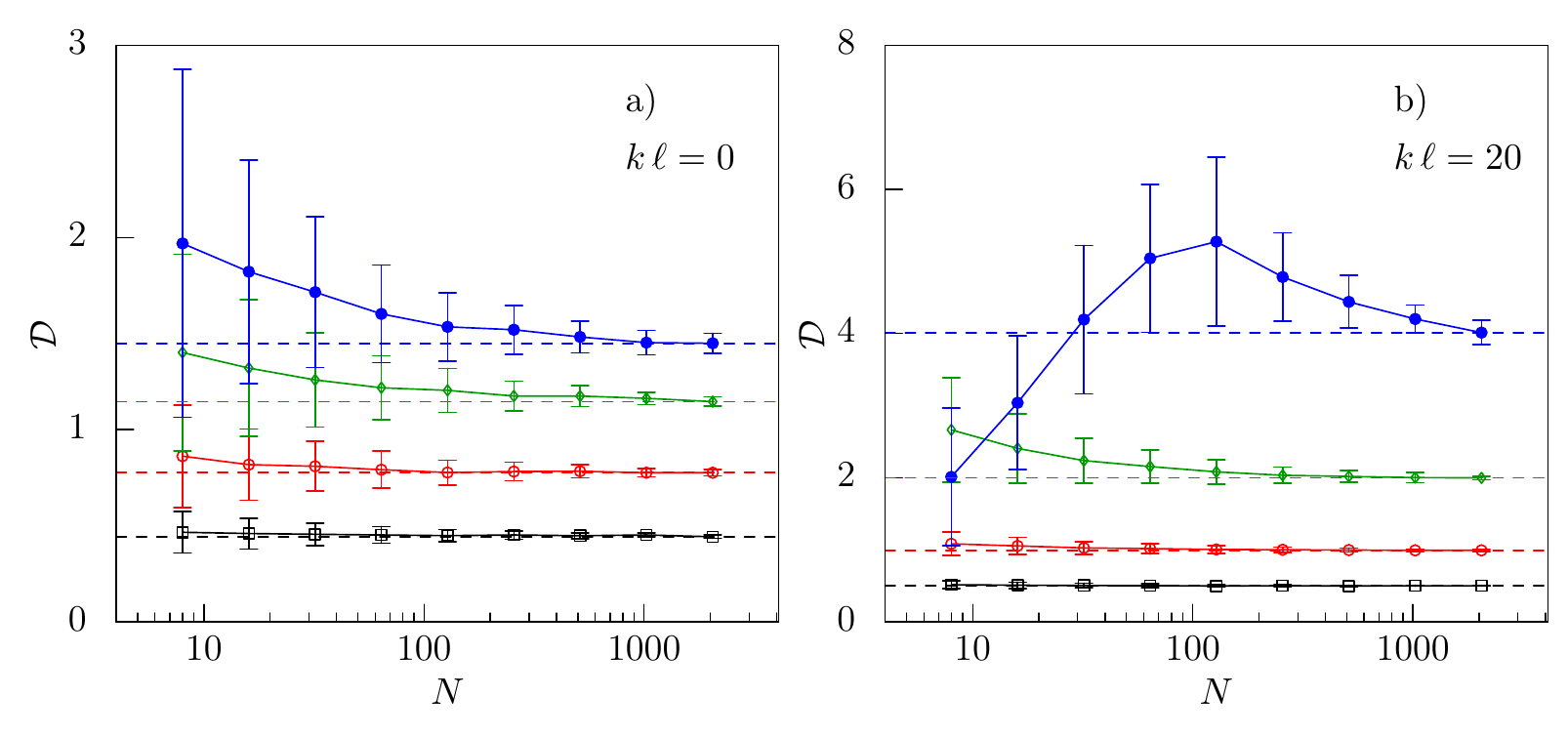}
\caption{Variation of the optical density $\OD=\ln|{\cal T}|^{-2}$ calculated from (\ref{eq:relation_T_Pi}) as function of the number of atoms $N$, for $\ell=0$ (a) and $\ell=20\,k^{-1}$ (b), and for 4 values of the density: $\sigma_0 \ntwoD=0.5$ (black), $1$ (red), 2 (green) and 4 (blue).
The bars indicate the standard deviations. The dotted lines give the value obtained for our largest value of $N$ ($N=2048$). The results have been obtained at resonance ($\Delta=0$).
}
\label{fig:extensivity}
\end{center}
\end{figure}

We start our study by testing the minimal atom number that is necessary to obtain a faithful estimate of the optical density. We choose a given value of $\ncol=N/\pi R^2$ and we investigate how $\OD$ depends on $N$ either for a strictly 2D gas ($\ell=0$) or for a gas extending significantly along the third direction ($\ell=20\,k^{-1}$). We consider a resonant probe for this study ($\Delta=0$). We vary $N$ by multiplicative steps of 2, from $N=8$ up to $N=2048$ and we determine how large $N$ must be so that $\OD$ is a function of $\ncol$ only. 

The results are shown in Fig.\,\ref{fig:extensivity}a and Fig.\,\ref{fig:extensivity}b, where we plot $\OD$ as a function of $N$. We perform this study for four values of the density $\ncol$, corresponding to $\sigma_0 \ncol=0.5, 1,2$ and 4. Let us consider first the smallest value $\sigma_0\ncol=0.5$. For each value of $N$ we perform a number of draws that is sufficient to bring the standard error below $2\times 10^{-3}$ and we find that the calculated optical density is independent of $N$ (within standard error) already for $N\gtrsim 100$, for both values of $\ell$. Consider now our largest value $\sigma_0\ncol=4$; for a strictly 2D gas ($\ell=0$), $\OD$ reaches an approximately constant value independent of $N$ for $N \gtrsim 1000$. For $\sigma_0 \ncol=4$ and a relatively thick gas ($\ell=20\,k^{-1}$, blue squares in Fig.\,\ref{fig:extensivity}b), reaching the thermodynamic limit is more problematic since there is still a clear difference between the results obtained with 1024 and 2048 atoms. This situation thus corresponds to the limit of validity of our numerical results. In the remaining part of the paper we will show only results obtained with $N=2048$ atoms for column densities not exceeding $\sigma_0 \ncol=4$. The number of independent draws of the atomic positions (at least 8) is chosen such that the standard error for each data point is below 2\%.

\subsection{Measured optical density vs. Beer--Lambert prediction}
\label{subsec:OD_vs_ODBL}

\begin{figure}[t]
\begin{center}
\includegraphics{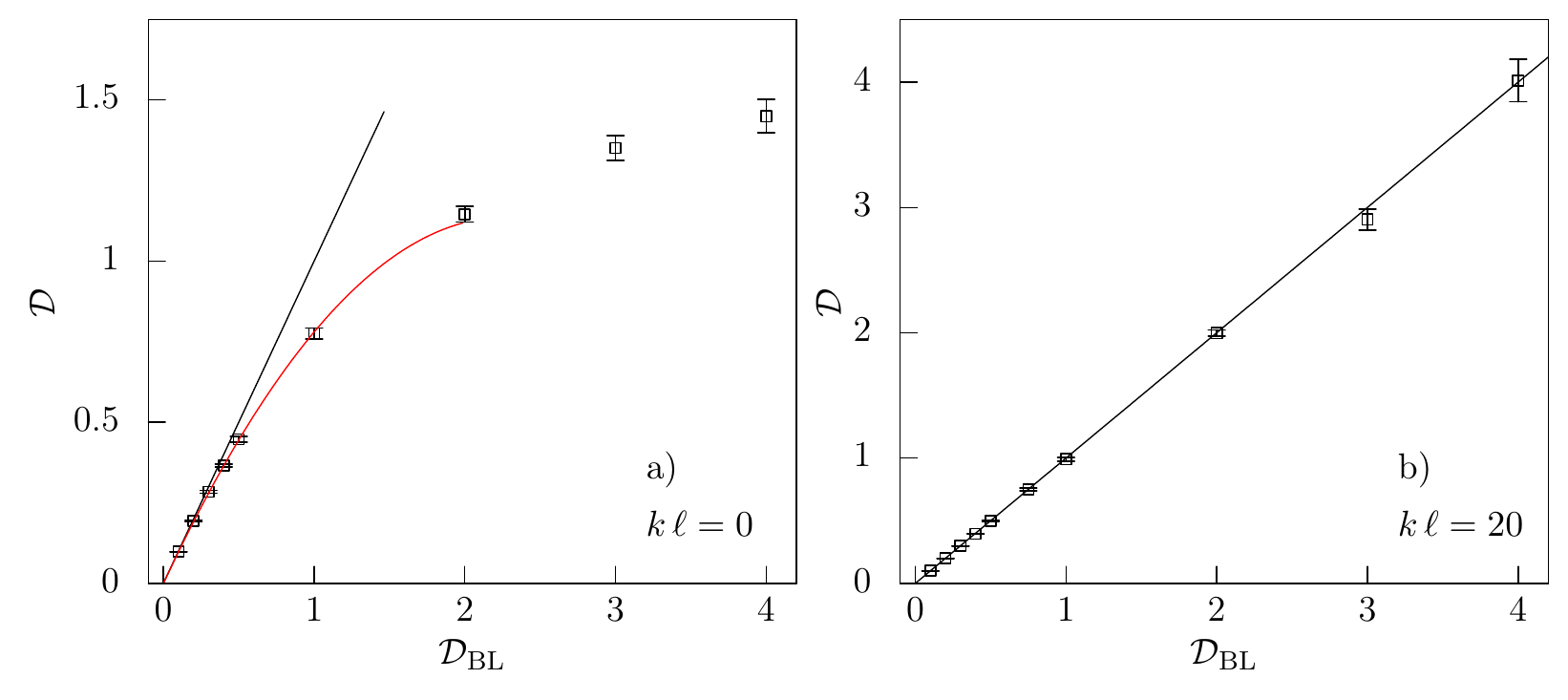}
\caption{Variations of the optical density $\OD$ as function of the Beer--Lambert prediction $\ODBL$ for $\ell=0$ (a) and $\ell=20\,k^{-1}$ (b). The black dotted line is the straight line of slope 1. In (a) the continuous red line is a quadratic fit $\OD = \ODBL\left(1-\mu\, \ODBL\right)$ with $\mu=0.22$ to the data points with $\ODBL \leq 1$. The calculations are done for $N=2048$, $\Delta=0$ and the bars indicate standard deviations.}
\label{fig:OD_ODnaive}
\end{center}
\end{figure}
 
We now investigate the variation of the optical density $\OD=\ln|{\cal T}|^{-2}$ as function of the column density of the sample $\ncol$, or equivalently of the Beer--Lambert prediction $\ODBL=\ncol\sigma$.  We suppose in this section that the probe beam is resonant ($\Delta=0$), and we address the cases of a strictly 2D gas ($\ell=0$) and a thick slab  ($\ell=20\,k^{-1}$).

Consider first the case of a strictly 2D case, $\ell=0$, leading to the results shown in Fig.\,\ref{fig:OD_ODnaive}a. We see that $\OD$ differs significantly ($\sim\,25$\%) from  $\ODBL$ already for $\ODBL$ around 1. A quadratic fit to the calculated variation of $\OD$ for  $\sigma_0 \ntwoD < 1$ (continuous red line) gives
\begin{equation}
\OD \approx \ODBL\left(1-0.22\, \ODBL \right) .
\label{eq:OD_close_to_0}
\end{equation} 
The discrepancy between $\OD$ and $\ODBL$ increases when the density increases: for $\ODBL=4$, the calculated $\OD$  is only $\approx 1.4$. For such a large density the average distance between nearest neighbours is $\approx k^{-1}$ and the energy shifts due to the dipole-dipole interactions are comparable to or larger than the linewidth $\Gamma$. The atomic medium is then much less opaque to a resonant probe beam than in the absence of dipole-dipole coupling.    

Consider now the case of a thick sample, $\ell=20\,k^{-1}$ (Fig.\,\ref{fig:OD_ODnaive}b). The calculated optical density is then very close to the Beer--Lambert prediction over the whole range that we studied.  This means that in our chosen range of optical densities, the mean-field approximation leading to $\ODBL$ is satisfactory as soon as the sample thickness exceeds a few optical wavelengths $\lambda=2\pi/k$.

\begin{figure}[tb]
\begin{center}
\includegraphics{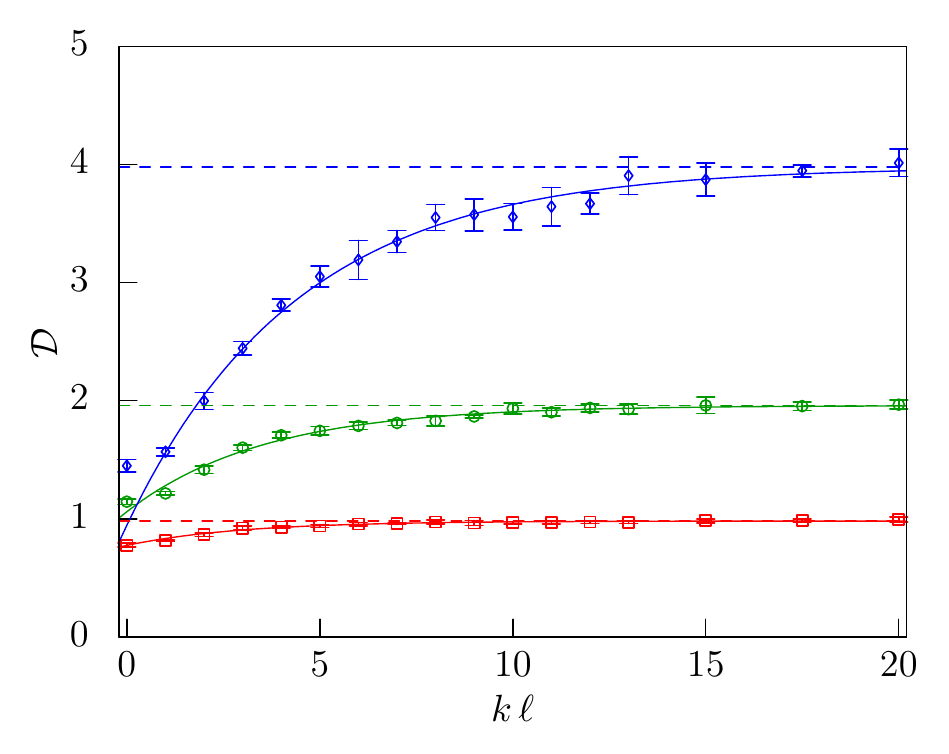}
\caption{Variation of $\OD$ with the thickness $\ell$ of the gas for various column densities corresponding to $\ODBL=1$ (red), 2 (green), 4 (blue). The continuous lines are exponential fits to the data. The dotted lines give the Beer--Lambert result. The calculations are done for $N=2048$, $\Delta=0$ and the bars indicate standard deviations. }
\label{fig:epaisseur}
\end{center}
\end{figure}

It is interesting to characterize how the optical density evolves from the value for a strictly 2D gas to the expected value from the Beer--Lambert law $\ODBL$ when the thickness of the gas increases. We show in Fig. \ref{fig:epaisseur}  the variation of  ${\cal D}$ as function of $\ell$ for three values of the column density  corresponding to $\ODBL=1,2$ and 4. An exponential fit ${\cal D}=\alpha + \beta \exp(-\ell/\ell_c)$ to these data for $2\,k^{-1}\leq\ell \leq 20\,k^{-1}$ gives a good account of the observed variation over this range, and it provides the characteristic thickness $\ell_c$ needed to recover the Beer--Lambert law. We find that $\ell_c\approx 3.0\,k^{-1}$ for $\ODBL=1$, $\ell_c\approx 3.5\,k^{-1}$ for $\ODBL=2$,  and $\ell_c\approx 4.4\,k^{-1}$ for $\ODBL=4$.

\paragraph{Remark.} For the largest value of the column density considered here ($\ncol \sigma_0=4$) we find that $\OD$ increases slightly above the value $\ODBL$ when $\ell$ is chosen larger than $20\,k^{-1}$ (upper value considered in Fig. \ref{fig:epaisseur}).  We believe that this is a consequence of the edge terms that we neglected when approximating  Eq. (\ref{eq:Ex_sur_EL}) by Eq. (\ref{eq:relation_T_Pi}). These terms become significant for $\ODBL=4$ because for our atom number $N=2048$, the sample radius $R\approx 55\,k^{-1}$ is then not very large compared to its thickness for $\ell \gtrsim 20\,k^{-1}$. In order to check this assumption, we also calculated numerically the result of Eq. (\ref{eq:Ex_sur_EL})  (instead of Eq. (\ref{eq:relation_T_Pi})) for practical values of the parameters (position and radius) of the lens represented in Fig. \ref{fig:experiment}a. The results give again $\OD\approx\ODBL$, but now with $\OD$ remaining below $\ODBL$. Since our emphasis in this paper is rather put on the 2D case, we will not explore this aspect further here.

\subsection{Absorption line shape}

\begin{figure}[t]
\begin{center}
\includegraphics{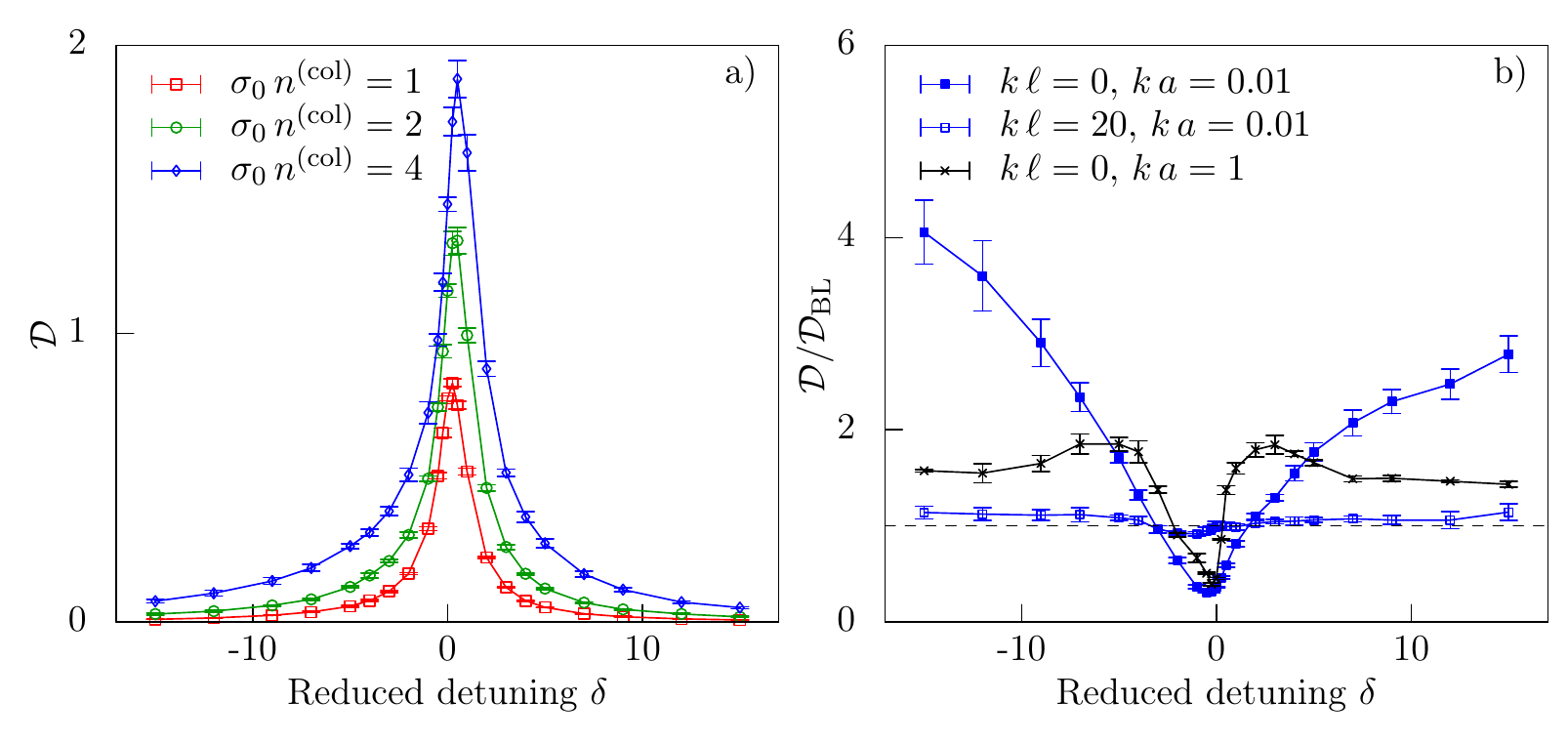}
\caption{(a) Variation of $\OD$ with the reduced detuning $\delta=2\Delta/\Gamma$ of the probe laser in the case of a 2D gas ($\ell=0$), for three values of the column density  $\sigma_0 \ncol=1$ (red), 2 (green), and 4 (blue).   (b) Blue full squares: same data as in (a), now plotted for $\OD/\ODBL$ as function of $\delta$. Blue open squares: $\OD/\ODBL$ for a thick gas ($\ell=20\,k^{-1}$). Black stars: $\OD/\ODBL$ for a 2D gas ($\ell=0$) and a large exclusion region around each atom ($a=k^{-1}$). All data in (b) correspond to $\sigma_0 \ntwoD=4$. The calculations are done with $N=2048$ atoms and the bars indicate standard deviations.}
\label{fig:detuning}
\end{center}
\end{figure}

Resonant van der Waals interactions manifest themselves not only in the reduction of the optical density at resonance but also in the overall line shape of the absorption profile. To investigate this problem we 
have studied the variations of $\OD$ with the detuning of the probe laser. We show in Fig. \ref{fig:detuning}a the results for a strictly 2D gas ($\ell=0$) for $\ncol\sigma_0=1,2$ and 4.  Several features show up in this series of plots. First we note a blue shift of the resonance, which increases with $\ntwoD$ and reaches $\Delta \approx \Gamma/4$ for $\sigma_0 \ntwoD=4$. We also note a slight broadening of the central part of the absorption line, since the full-width at half maximum, which is equal to $\Gamma$ for an isolated atom,  is $\simeq 1.3 \Gamma$ for $\ncol \sigma_0=4$. Finally we note the emergence of large, non-symmetric wings in the absorption profile. This asymmetry is made more visible in Fig. \ref{fig:detuning}b, where we show with full blue squares the same data as in Fig. \ref{fig:detuning}a for $\ncol\sigma_0=4$, but now plotting $\OD/\ODBL$ as function of $\delta$. For a detuning $\delta=\pm 15$, the calculated optical density exceeds the Beer--Lambert prediction by a factor 4.1 (resp. 2.8) on the red (resp. blue) side.

In order to get a better understanding of these various features, we give in Fig. \ref{fig:detuning}b two additional results. On the one hand we plot with empty blue squares the variations of $\OD/\ODBL$ for a thick gas ($\ell=20\,k^{-1}$) with the same column density $\ncol\sigma_0=4$. There are still some differences between $\OD$ and $\ODBL$ in this case, as already pointed out in \cite{Sokolov:2009}, but they are much smaller than in the $\ell=0$ case. This indicates that the strong deviations with respect to the Beer--Lambert law that we observe in Fig. \ref{fig:detuning}a  are specific 2D features. On the other hand we plot with black stars the variations of $\OD/\ODBL$ for a 2D gas ($\ell=0$) in which we artificially increased the exclusion radius around each atom up to $a=k^{-1}$ instead of $a=0.01\,k^{-1}$ (blue full squares) for the other results in this paper. This procedure, which was suggested to us by Robin Kaiser, allows one to discriminate between effects due to isolated pairs of closely spaced atoms, and many-body
features resulting from multiple scattering of photons among larger clusters of atoms. The comparison of the results obtained for $a=0.01\,k^{-1}$ and $a=k^{-1}$ suggests that the blue shift of the resonance line, which is present in both cases, is a many-body phenomenon, whereas the large amplitude wings with a blue-red asymmetry, which occurs only for $a=0.01\,k^{-1}$, is rather an effect of close pairs.

This asymmetry in the wings of the absorption line in a 2D gas can actually be understood in a semi-quantitative manner by a simple reasoning. We recall that for two atoms at a distance $r \ll k^{-1}$, the levels involving one ground and one excited atom have an energy (real part of the eigenvalues of $H_{\rm eff}$) that is displaced by $\sim \pm \hbar \Gamma  /(kr)^3$. A given detuning $\delta$ can thus be associated to a distance $r$ between the two members of a pair that will resonantly absorb the light. To be more specific let us consider a pair of atoms with $kr \ll 1$, and suppose for simplicity that it is aligned either along the polarization axis of the light ($x$) or perpendicularly to the axis ($y$). In both cases the excited state of the pair that is coupled to the laser is the symmetric combination $ (|1:g;2:e_x\rangle+|1:e_x;2:g\rangle)/\sqrt 2$. If the pair is aligned along the $x$ axis, this state has an energy $\hbar \omega_0-3 \hbar \Gamma/2(kr)^3$, hence it is resonant with red detuned light such that $\delta=-3/(kr)^3$. If the pair axis is perpendicular to $x$, the state written above has an energy $\hbar \omega_0+3 \hbar \Gamma/4(kr)^3$, hence it is resonant with blue detuned light such that $\delta=3/2(kr)^3$. This clearly leads to an asymmetry between red and blue detuning; indeed the pair distance $r$ needed for ensuring resonance  for a given $\delta>0$, $r_{\rm blue}=(3/2|\delta|)^{1/3}$, is smaller than the value $r_{\rm red}=(3/|\delta|)^{1/3}$ for the opposite value $-\delta$. Since the probability density for the pair distance is ${\cal P}(r)\propto r$ in 2D for randomly drawn positions, we expect the absorption signal to be stronger for $-\delta$ than for $+\delta$. In a 3D geometry the variation of the probability density with $r$ is even stronger (${\cal P}(r)\propto r^2$), but it is compensated by the fact that the probability of occurrence of pairs that are resonant with blue detuned light is dimensionally increased. For example in our simplified modelling where the pair axis is aligned with the references axes, a given pair will be resonant with blue detuned light in 2/3 of the cases (axis along $y$ or $z$) and resonant with red detuned light only in 1/3 of the cases (axis along $x$). This explains why the asymmetry of the absorption profile is much reduced for a 3D gas in comparison to the 2D case.


\section{Summary}
\label{sec:summary}

We have presented in this paper a detailed analysis of the scattering of light by a disordered distribution of atoms in a quasi-two dimensional geometry. The particles were treated as fixed scatterers and their internal structure was modeled as a two-level system, with a $J=0$ ground state and a $J=1$ excited state. In spite of these simplifying assumptions the general trend of our results is in good agreement with the experimental finding of \cite{Rath:2010}, where a variation of the measured optical density similar to that of Fig.~\ref{fig:OD_ODnaive} was  measured. 

Several improvements in our modeling can be considered in order to reach a quantitative agreement with theory and experiment. The first one is to include the relatively complex atomic structure of the alkali-metal species used in practice, with a multiply degenerate ground state; this could be done following the lines of  \cite{Jonckheere:2000,Muller:2002}. A second  improvement consists in taking into account the atomic motion. This is in principle a formidable task, because it leads to a spectacular increase in the dimension of the relevant Hilbert space. This addition can however be performed in practice in some limiting cases, for example if one assumes that the particles are tightly bound in a lattice \cite{Antezza:2009,Antezza:2009b}. When the atom-light interaction is used only to probe the spatial atomic distribution of the gas, neglecting the particle motion should not be a major problem. Indeed the duration of the light pulse is quite short ($\sim 10$ microseconds only). Each atom scatters only a few photons in this time interval and its displacement is then smaller than the mean interatomic spacing for the spatial densities encountered in practice. The acceleration of the atoms under the effect of resonant van der Waals interaction should also have a minor effect under relevant experimental conditions. Finally another aspect that could be valuably studied is the interaction of the gas with an intense laser beam \cite{Reinaudi:2007}. One could thus validate the intuitive idea that saturation phenomena reduce the effects of resonant van der Waals interactions \cite{Hung:2011,Yefsah:2011}, and are thus helpful to provide a faithful estimate of the atomic density from the light absorption signal.

\ack
We thank I. Carusotto, Y. Castin, K. G\"unter, M. Holzmann, R. Kaiser, W. Krauth and S.P. Rath for helpful discussions and comments. This work is supported by IFRAF and ANR (project  BOFL).

\section*{References}
\bibliographystyle{unsrt.bst}

\begin{thebibliography}{10}

\bibitem{Dalfovo:1999}
F.~S. Dalfovo, L.~P. Pitaevkii, S.~Stringari, and S.~Giorgini.
\newblock Theory of {B}ose--{E}instein condensation in trapped gases.
\newblock {\em Rev. Mod. Phys.}, 71:463, 1999.

\bibitem{Lewenstein:2007}
M.~Lewenstein, A.~Sanpera, V.~Ahufinger, B.~Damski, A.~Sen De, and U.~Sen.
\newblock Ultracold atomic gases in optical lattices: mimicking condensed
  matter physics and beyond.
\newblock {\em Adv. Phys.}, 56(2):243--379, 2007.

\bibitem{Bloch:2008}
I.~Bloch, J.~Dalibard, and W.~Zwerger.
\newblock Many-body physics with ultracold gases.
\newblock {\em Rev. Mod. Phys}, 80(3):885, 2008.

\bibitem{Giorgini:2008}
S.~Giorgini, L.~P. Pitaevskii, and S.~Stringari.
\newblock Theory of ultracold atomic {F}ermi gases.
\newblock {\em Rev. Mod. Phys.}, 80:1215--1274, 2008.

\bibitem{Ketterle:1999b}
W.~Ketterle, D.~S. Durfee, and D.~M. Stamper-Kurn.
\newblock Making, probing and understanding {B}ose--{E}instein condensates.
\newblock In M.~Inguscio, S.~Stringari, and C.E. Wieman, editors, {\em
  {B}ose--{E}instein condensation in atomic gases}, Proceedings of the
  International School of Physics Enrico Fermi, Course CXL, page~67, Amsterdam,
  1999. IOS Press.

\bibitem{Ho:2009}
T.-L. Ho and Qi~Zhou.
\newblock Obtaining the phase diagram and thermodynamic quantities of bulk
  systems from the densities of trapped gases.
\newblock {\em Nature Physics}, 6:131, 2009.

\bibitem{Rath:2010}
S.~P. Rath, T.~Yefsah, K.~J. G\"unter, M.~Cheneau, R.~Desbuquois, M.~Holzmann,
  W.~Krauth, and J.~Dalibard.
\newblock Equilibrium state of a trapped two-dimensional {B}ose gas.
\newblock {\em Phys. Rev. A}, 82:013609, 2010.

\bibitem{Yefsah:2011}
T.~Yefsah, R.~Desbuquois, L.~Chomaz, K.~J. G\"unter, and J.~Dalibard.
\newblock Exploring the thermodynamics of a two-dimensional {B}ose gas.
\newblock {\em Phys. Rev. Lett.}, 107:130401, 2011.

\bibitem{Labeyrie:1999}
G.~Labeyrie, F.~de~Tomasi, J.-C. Bernard, C.~A. M\"uller, C.~Miniatura, and
  R.~Kaiser.
\newblock Coherent backscattering of light by cold atoms.
\newblock {\em Phys. Rev. Lett.}, 83:5266--5269, 1999.

\bibitem{Labeyrie:2003}
G.~Labeyrie, E.~Vaujour, C.~A. M\"uller, D.~Delande, C.~Miniatura,
  D.~Wilkowski, and R.~Kaiser.
\newblock Slow diffusion of light in a cold atomic cloud.
\newblock {\em Phys. Rev. Lett.}, 91:223904, 2003.

\bibitem{Akkermans:2007}
E.~Akkermans and G.~Montambaux.
\newblock {\em Mesoscopic {P}hysics of {E}lectrons and {P}hotons}.
\newblock Cambridge University Press, Cambridge, England, 2007.

\bibitem{Morice:1995}
O.~Morice, Y.~Castin, and J.~Dalibard.
\newblock Refractive index of a dilute {B}ose gas.
\newblock {\em Phys. Rev. A}, 51:3896, 1995.

\bibitem{Pinheiro:2004}
F.~A. Pinheiro, M.~Rusek, A.~Orlowski, and B.~A. van Tiggelen.
\newblock Probing {A}nderson localization of light via decay rate statistics.
\newblock {\em Phys. Rev. E}, 69:026605, 2004.

\bibitem{Gero:2007}
A.~Gero and E.~Akkermans.
\newblock Superradiance and multiple scattering of photons in atomic gases.
\newblock {\em Phys. Rev. A}, 75:053413, 2007.

\bibitem{Svidzinsky:2008}
A.~A. Svidzinsky, J.-T. Chang, and M.~O. Scully.
\newblock Dynamical evolution of correlated spontaneous emission of a single
  photon from a uniformly excited cloud of ${N}$ atoms.
\newblock {\em Phys. Rev. Lett.}, 100:160504, 2008.

\bibitem{Akkermans:2008}
E.~Akkermans, A.~Gero, and R.~Kaiser.
\newblock Photon localization and {D}icke superradiance in atomic gases.
\newblock {\em Phys. Rev. Lett.}, 101:103602, 2008.

\bibitem{Sokolov:2009}
I.~M. Sokolov, M.~D. Kupriyanova, D.~V. Kupriyanov, and M.~D. Havey.
\newblock Light scattering from a dense and ultracold atomic gas.
\newblock {\em Phys. Rev. A}, 79:053405, 2009.

\bibitem{Scully:2009}
M.~O. Scully.
\newblock Collective {L}amb shift in single photon {D}icke superradiance.
\newblock {\em Phys. Rev. Lett.}, 102:143601, 2009.

\bibitem{Goetschy:2011b}
A.~Goetschy and S.~E. Skipetrov.
\newblock Euclidean matrix theory of random lasing in a cloud of cold atoms.
\newblock {\em EPL (Europhysics Letters)}, 96:34005, 2011.

\bibitem{cohe92}
C.~Cohen-Tannoudji, J.~Dupont-Roc, and G.~Grynberg.
\newblock {\em Atom-Photon Interactions}.
\newblock Wiley, New York, 1992.

\bibitem{Jonckheere:2000}
T.~Jonckheere, C.~A. M\"uller, R.~Kaiser, C.~Miniatura, and D.~Delande.
\newblock Multiple scattering of light by atoms in the weak localization
  regime.
\newblock {\em Phys. Rev. Lett.}, 85:4269--4272, 2000.

\bibitem{Muller:2002}
C.~A. M\"uller and C.~Miniatura.
\newblock Multiple scattering of light by atoms with internal degeneracy.
\newblock {\em Journal of Physics A: Mathematical and General}, 35(47):10163,
  2002.

\bibitem{Cohen:1989}
C.~Cohen-Tannoudji, J.~Dupont-Roc, and G.~Grynberg.
\newblock {\em Photons and {A}toms--{I}ntroduction to {Q}uantum
  {E}lectrodynamics}.
\newblock Wiley, New-York, 1989.

\bibitem{Messiah:scattering}
A.~Messiah.
\newblock {\em Quantum {M}echanics, Chapter XIX}, volume~II.
\newblock North-Holland Publishing Company, Amsterdam, 1961.

\bibitem{Morice:1995b}
O.~Morice.
\newblock {\em Atomes refroidis par laser : du refroidissement sub-recul \`a la
  recherche d'effets quantiques collectifs}.
\newblock PhD thesis, Universit{\'e} Pierre et Marie Curie, Paris,
  http://tel.archives-ouvertes.fr/docs/00/06/13/10/PDF/1995MORICE.pdf, 1995.

\bibitem{Jackson:book}
J.~D. Jackson.
\newblock {\em Classical {E}lectrodynamics}.
\newblock John Wiley, New York, 1998.

\bibitem{Mezard:1999}
A.~Zee M.~M\'ezard, G.~Parisi.
\newblock Spectra of euclidean random matrices.
\newblock {\em Nuclear Physics B}, 559:689, 1999.

\bibitem{Rusek:2000}
M.~Rusek, J.~Mostowski, and A.~Or\l{}owski.
\newblock Random green matrices: From proximity resonances to {A}nderson
  localization.
\newblock {\em Phys. Rev. A}, 61:022704, 2000.

\bibitem{Skipetrov:2011}
S.~E. Skipetrov and A.~Goetschy.
\newblock Eigenvalue distributions of large {E}uclidean random matrices for
  waves in random media.
\newblock {\em Journal of Physics A: Mathematical and Theoretical},
  44(6):065102, 2011.

\bibitem{Goetschy:2011}
A.~Goetschy and S.~E. Skipetrov.
\newblock Non-{H}ermitian {E}uclidean random matrix theory.
\newblock {\em Phys. Rev. E}, 84:011150, 2011.

\bibitem{Foldy:1945}
L.~L. Foldy.
\newblock The multiple scattering of waves. i. general theory of isotropic
  scattering by randomly distributed scatterers.
\newblock {\em Phys. Rev.}, 67:107--119, 1945.

\bibitem{Stephen:1964}
M.~J. Stephen.
\newblock First-order dispersion forces.
\newblock {\em J. Chem. Phys.}, 40:669, 1964.

\bibitem{Hutchinson:1964}
D.~A. Hutchinson and H.~F. Hameka.
\newblock Interaction effects on lifetimes of atomic excitations.
\newblock {\em J. Chem. Phys.}, 41:2006, 1964.

\bibitem{Dicke:1954}
R.~H. Dicke.
\newblock Coherence in spontaneous radiation processes.
\newblock {\em Phys. Rev.}, 93:99, 1954.

\bibitem{Prokofev:2001}
N.~V. Prokof'ev, O.~Ruebenacker, and B.~V. Svistunov.
\newblock Critical point of a weakly interacting two-dimensional {B}ose gas.
\newblock {\em Phys. Rev. Lett.}, 87:270402, 2001.

\bibitem{Antezza:2009}
M.~Antezza and Y.~Castin.
\newblock Spectrum of light in a quantum fluctuating periodic structure.
\newblock {\em Phys. Rev. Lett.}, 103:123903, 2009.

\bibitem{Antezza:2009b}
M.~Antezza and Y.~Castin.
\newblock {F}ano-{H}opfield model and photonic band gaps for an arbitrary
  atomic lattice.
\newblock {\em Phys. Rev. A}, 80:013816, 2009.

\bibitem{Reinaudi:2007}
G.~Reinaudi, T.~Lahaye, Z.~Wang, and D.~Gu\'ery-Odelin.
\newblock Strong saturation absorption immaging of dense clouds of ultracold
  atoms.
\newblock {\em Opt. Lett.}, 32:3143, 2007.

\bibitem{Hung:2011}
Chen-Lung Hung, Xibo Zhang, Nathan Gemelke, and Cheng Chin.
\newblock Observation of scale invariance and universality in two-dimensional
  {B}ose gases.
\newblock {\em Nature}, 470:236, 2011.

\end{thebibliography}

\end{document}